\documentclass[prc,twocolumn,floatfix,groupedaddress,nofootinbib,showpacs,preprintnumbers,
amsmath,amssymb,amsfonts,superscriptaddress,widetable] {revtex4-1}
\usepackage{graphicx}
\usepackage{dcolumn}
\usepackage{bm}
\usepackage{amsmath}
\usepackage{mathrsfs}
\usepackage{url}
\usepackage{relsize}
\def\expo{{\mathlarger{e}}}
\usepackage{xcolor}

\begin{document}

\title{The Dirac oscillator: an alternative basis for nuclear structure calculations}

\author{Junjie Yang and J. Piekarewicz}
\affiliation{%
 Department of Physics, Florida State University, Tallahassee, FL 32306, USA}%

\date{\today}

\begin{abstract}

\noindent\textbf{Background:} The isotropic harmonic oscillator supplemented by
a strong spin-orbit interaction has been the cornerstone of nuclear structure since
its inception more than seven decades ago. In this paper we introduce---or rather
re-introduced---the ``Dirac Oscillator", a fully relativistic basis that has all the 
desired attributes of the ordinary harmonic oscillator while naturally incorporating
a strong spin-orbit coupling.
\smallskip

\noindent\textbf{Purpose:} To assess---to our knowledge for the first time---the power 
and flexibility of the Dirac Oscillator basis in the solution of nuclear structure problems 
within the framework of covariant density functional theory.
\smallskip

\noindent\textbf{Methods:} Self-consistent calculations of binding energies and
ground-state densities for a selected set of doubly-magic magic are performed 
using the Dirac-oscillator basis and are then compared against results obtained 
with the often-used Runge-Kutta method.
\smallskip

\noindent\textbf{Results:} Results obtained using the Dirac oscillator basis 
reproduced with high accuracy those derived using the Runge-Kutta method
and suggest a clear path for a generalization to systems with axial symmetry.
\smallskip

\noindent\textbf{Conclusions:} Although the three-dimensional harmonic oscillator 
with spin-orbit corrections has been the staple of the nuclear shell model since the 
beginning, the Dirac oscillator is practically unknown among the nuclear 
physics community. In this paper we illustrate the power and flexibility of the
Dirac oscillator and suggest extensions to the study of systems without spherical 
symmetry as required in constrained calculations of nuclear excitations.

\end{abstract}

\maketitle


\section{\label{sec:intro}Introduction}

The isotropic, three-dimensional harmonic oscillator has played a critical role in the 
development of the nuclear-shell model since its inception in the late 1940s. Indeed, 
Haxel, Jensen, and Suess start their 1949 seminal paper with: ``A simple explanation 
of the \emph{magic numbers} 14, 28, 50, 82, 126 follows at once from the oscillator 
model of the nucleus, if one assumes that the spin-orbit coupling in the Yukawa field 
theory of nuclear forces leads to a strong splitting ..."\,\cite{Haxel:1949}. Independently
and just two weeks later, Maria Goeppert-Mayer provides detailed evidence supporting 
the emergence of magic numbers as a consequence of a strong spin-orbit 
coupling\,\cite{Mayer:1949}. In that paper as well as in her 1963 Nobel lectures, 
Goeppert-Mayer credits Enrico Fermi with a profound insight: ``One day as Fermi 
was leaving my office he asked, {\sl Is there any indication of spin-orbit coupling?"}
\cite{Mayer:1963}. 

Fast forward to today and despite remarkable advances in both 
refining the underlying interaction and perfecting the many-body methods, the 
nuclear-structure community continues to rely heavily on the three-dimensional 
harmonic oscillator with spin-orbit corrections as a convenient and flexible 
single-particle basis; see Refs.\,\cite{Brown:2001zz,Dickhoff:2004xx,Roth:2010bm,
Kortelainen:2010hv,Hagen:2013nca,Barrett:2013nh} and references contained therein 
for a representative set of modern approaches to nuclear structure. Among the reasons
that the harmonic oscillator continues to be heavily used is because it permits a clean 
separation of the center-of-mass motion and has convenient analytic properties;  e.g., it 
has the same functional form in momentum space as in configuration space.

Although highly successful, relativistic approaches to nuclear structure are more limited
in scope and generally fall under the single rubric of covariant density functional 
theory (DFT)\,\cite{Serot:1984ey,Meng:2016}. The aim of covariant DFT is to build 
high-quality functionals that yield an accurate description of the properties of finite nuclei, 
generate an equation of state that is consistent with known neutron-star properties, while 
providing a Lorentz covariant extrapolation to dense matter. However, given that the model 
parameters underlying the DFT cannot be computed from first principles, their values must
be calibrated from a suitable set of experimental data; see Ref.\,\cite{Chen:2014sca} and 
references contained therein. From such an optimally calibrated density functional, one 
derives the corresponding Kohn-Sham (or mean-field) equations which are then solved 
self-consistently\,\cite{Horowitz:1981xw}. In particular, the nucleon field satisfies a Dirac
equation in the presence of strong Lorentz scalar and vector potentials that naturally 
lead to a very strong spin-orbit splitting. Given that the Dirac equation for spherically
symmetric potentials separates into a set of two coupled differential equations in the
radial coordinate, one often solves this set of equations using a conventional 
Runge-Kutta algorithm\,\cite{Horowitz:1981xw,Serot:1984ey,Todd:2003xs}. Alternatively, 
one may solve the Dirac equation as a matrix-diagonalization problem by expanding both 
the upper and lower components of the Dirac spinor in a {\sl non-relativistic} harmonic oscillator 
basis; see for example Ref.\,\cite{Meng:2016,Vretenar:2005zz} and references contained 
therein. There is, however, a more natural alternative.

Back in 1989 Moshinsky and Szczepaniak modified the free Dirac equation---already 
linear in the momentum---to one that would be linear in \emph{both} the coordinates 
and the momenta of the particle\,\cite{Moshinsky:1989}; see also a much earlier paper 
by It{\^o}, Mori, and Carriere\,\cite{Ito:1967}. By doing so, they introduced the ``The Dirac 
Oscillator", a problem that can be solved exactly as both upper and lower components 
of the Dirac equation satisfy a \emph{non-relativistic} harmonic oscillator problem with a 
strong spin-orbit coupling term. Although the paper has generated considerable interest in 
certain fields, we find surprising that it has not generated as much excitement in the nuclear 
structure community, given the prominent role that the harmonic oscillator potential 
supplemented by a strong spin-orbit coupling has played in nuclear physics for so 
many decades. As we show below, however, the spectrum of the Dirac oscillator is 
quite different from that of the ordinary non-relativistic harmonic 
oscillator\,\cite{Quesne:1990}.

The aim of this paper is to introduce and illustrate the value of the Dirac oscillator as 
a complete basis for the solution of the relativistic Kohn-Sham equations. Comparisons 
will be made against solutions obtained using the standard Runge-Kutta method. We
will also argue in favor of the Dirac oscillator basis over the Runge-Kutta method for 
the treatment of problems with axial symmetry. The rest of the paper is 
organized as follows. In Sec.\,\ref{sec:form} we briefly review the basic ideas behind
covariant DFT. Later on in the section we introduce the Dirac oscillator and obtained 
a system of equations that, as advertised, will become identical to the differential
equation satisfied by the ordinary harmonic oscillator supplemented by a strong 
spin-orbit term. In Sec.\,\ref{sec:result} we will show results obtained from matrix 
diagonalization in the Dirac oscillator basis and underscore the excellent agreement 
when compared against the Runge-Kutta method. Finally, in Sec.\,\ref{sec:conclusion}, 
we summarize our results and suggest other possible applications of the Dirac 
oscillator basis.

\section{\label{sec:form}Formalism}

\subsection{\label{subsec:rmf}Covariant Density Functional Theory}
In the context of covariant density functional theory, the basic degrees of freedom 
are nucleons (protons and neutrons) interacting via short-range nuclear interactions 
mediated by various ``mesons" and a long-range Coulomb interactions mediated by 
the photon. Since the early attempts at a relativistic description of the nuclear 
dynamics\,\cite{Johnson:1955zz,Duerr:1956zz,Miller:1972zza,Walecka:1974qa},
various refinements have been made by incorporating density-dependent interactions
via self and mixed non-linear meson couplings\,\cite{Boguta:1977xi,Serot:1984ey,
Mueller:1996pm,Lalazissis:1996rd,Lalazissis:2005de,Todd-Rutel:2005fa,Chen:2014sca,
Chen:2014mza}. 

In the framework of relativistic Kohn-Sham (or mean-field) theory, the nucleons satisfy a 
Dirac equation with strong scalar and (time-like) vector potentials that are generated by 
the various meson fields which, in the mean-field approximation, become classical fields. 
In turn, the classical meson fields satisfy Klein-Gordon equations containing both non-linear 
meson interactions and ground-state baryon densities as source terms. It is this interplay
that demands a self-consistent solution to the problem. 

For the purpose of this work, it is sufficient to know that the nucleons satisfy a Dirac 
equation with a DFT Hamiltonian containing scalar and time-like vector potentials. 
That is,
\begin{equation} 
 \hat{H}_{\rm DFT} = \bm{\alpha} \cdot \bm{p} + V(r) + \beta \Big(m + S(r)\Big),
 \label{eq: dirac hamiltonian}
\end{equation} 
where $S(r)$ and $V(r)$ are the scalar and vector potentials, respectively, 
$\bm{p}$ is the momentum operator, and $\bm{\alpha}$ and $\beta$ are the 
four $4\times4$ Dirac matrices defined as follows:
\begin{equation}
 \bm{\alpha}=\left(\begin{array}{cc}
 0 & \bm{\sigma}\\
 \bm{\sigma} & 0
 \end{array}\right) \quad{\rm and}\quad
 \beta=\left(\begin{array}{cc}
 1 & 0\\
 0 & -1
\end{array}\right).
\end{equation}
Note that we have adopted units in which $\hbar\!=\!c\!=\!1$. For a more comprehensive 
discussion of the covariant DFT formalism see Refs.\,\cite{Todd:2003xs,Chen:2014sca,Yang:2019fvs} 
and references contained therein. 

The eigenvalue problem associated with the Hamiltonian displayed in Eq.(\ref{eq: dirac hamiltonian}) 
can be solved in multiple ways. For example, the Dirac equation derived from the above Hamiltonian 
with spherically symmetric scalar and vector potentials results in a set of two coupled, first order, 
ordinary differential equations that may solved using the Runge-Kutta method. Alternatively, 
one can expand the Hamiltonian into a suitable basis and then extract the eigenvalues and 
corresponding eigenvectors by diagonalizing the resulting Hamiltonian 
matrix\,\cite{stoitsov1998solution,li2012deformed,Meng:2016,Vretenar:2005zz}. In this paper we 
illustrate---to our knowledge for the first time---how to perform the diagonalization of $\hat{H}_{\rm DFT}$
using the Dirac oscillator basis of Moshinsky and Szczepaniak\,\cite{Moshinsky:1989,MartinezRomero:1995}. 
These results will then be compared against those obtained using the Runge-Kutta method. In turn, the 
flexibility of the Dirac oscillator basis naturally suggests a generalization into the study of deformed nuclei. 

\subsection{\label{subsec:do}The Dirac Oscillator}

The Hamiltonian for the Dirac oscillator is obtained from the free Dirac Hamiltonian by demanding 
that: (a) the resulting Hamiltonian be linear in both the momenta $\bm{p}$ and coordinates $\bm{r}$ 
of the particle and (b) both upper and lower components of the Dirac spinor satisfy the conventional
harmonic oscillator differential equation. Moshinsky and Szczepaniak\,\cite{Moshinsky:1989} were 
able to satisfy both conditions by performing the following substitution:
\begin{align}
 \bm{p} \rightarrow \bm{p} - im\omega \beta \bm{r},
\end{align}
which in turn transformed the free Dirac Hamiltonian into the \emph{Dirac oscillator} Hamiltonian:
\begin{equation} 
 \hat{H} = \bm{\alpha}\!\cdot\!\Big(\bm{p}\!-\!im\omega \beta \bm{r}\Big)+ \beta m =
  \left(\begin{array}{cc}
      m & \bm{\sigma}\cdot\bm{\pi_{+}} \\
      \bm{\sigma}\cdot\bm{\pi_{-}} & -m 
\end{array}\right), 
 \label{eq:DirOscH}
\end{equation} 
where $\omega$ will be identified as the frequency of the harmonic oscillator and we have 
defined
\begin{equation} 
 \bm{\pi}_{\pm} = \bm{p} \pm im\omega \bm{r}.
 \label{Pis}
\end{equation} 

Given that the above Hamiltonian is rotationally invariant, the most general 
solution of the Dirac oscillator is of the form\,\cite{Sakurai:1967}:
\begin{equation}
\psi_{E\kappa m}(\bm{r})=
\left(\begin{array}{c}
\phi_{E\kappa m}(\bm{r})\\
\chi_{E\kappa m}(\bm{r})
\end{array}\right)
=
\left(\begin{array}{c}
\phantom{i}g_{E\kappa}(r)\left|+\kappa m\right\rangle \\
if_{E\kappa}(r)\left|-\kappa m\right\rangle
\end{array}\right),
\label{eq:general solution}
\end{equation}
where $g$ and $f$ are radial functions associated to the upper and lower components 
of the Dirac spinor, respectively. Note that the relative phase (``$i$") introduced above 
ensures that both $g$ and $f$ are real functions of $r$. Finally,  the quantum number 
$\kappa\!\ne\!0$ is a nonzero integer related to the spin-spherical harmonic resulting from 
the coupling of the orbital angular momentum to the intrinsic nucleon spin. That is, 
\begin{equation}
 |\kappa m \rangle\!\equiv\!| l\frac{1}{2}jm\rangle, 
 \hspace{5pt} j\!=\!|\kappa|\!-\!\frac{1}{2}, \hspace{5pt}
 l\!=\!\begin{cases}
   j\!+\!\frac{1}{2} & \text{if } \kappa\!>\!0, \\
   j\!-\!\frac{1}{2} & \text{if } \kappa\!<\!0.
     \end{cases}
 \label{DefKappa}    
\end{equation}
This indicates that whereas $g(r)$ and $f(r)$ share a common value of the total angular momentum $j$, 
they differ by one unit of orbital angular momentum. For example, assuming $\kappa\!=\!-1$ yields 
$j\!=\!1/2$, but an $s$-wave upper component with $l\!=\!0$ and a $p$-wave lower component with 
$l\!=\!1$. Using Eqs.(\ref{eq:DirOscH}) and (\ref{eq:general solution}) one derives the eigenvalue equation 
for the Dirac oscillator. That is,
\begin{subequations}
\begin{align}
  \Big(\bm{\sigma}\cdot\bm{\pi}_{+}\Big)\chi(\bm{r}) & = (E-m)\phi(\bm{r}), \\
  \Big(\bm{\sigma}\cdot\bm{\pi}_{-}\Big)\phi(\bm{r}) & = (E+m)\chi(\bm{r}). 
\end{align}
\label{DiracEqn0}
\end{subequations}

Although some details will be left to the appendix, we illustrate here some of the essential 
steps involved in obtaining an equivalent ``Schr\"odinger-like" equation for the upper component.
For positive energy states, it is convenient to express the lower component in terms of the upper
in order to avoid any potential singular denominator. From Eqs.(\ref{DiracEqn0}) we obtain 
\begin{subequations}
\begin{align}
  & \chi(\bm{r}) = \frac{(\bm{\sigma}\cdot\bm{\pi}_{-})}{(E+m)}\phi(\bm{r}), \label{DiracEqn1a} \\
  & (\bm{\sigma}\cdot\bm{\pi}_{+})(\bm{\sigma}\cdot\bm{\pi}_{-})\phi(\bm{r}) =(E^{2}-m^{2})\phi(\bm{r}).
\end{align}
\label{DiracEqn1}
\end{subequations}
As in the case of the free Dirac equation, the above result indicates how a set of coupled first-order
differential equations may be decoupled at the expense of generating a second order differential 
equation. After performing some standard spin algebra one obtains the following
Schr\"odinger-like equation for $\phi$:
\begin{equation} 
 \left(\frac{\bm{p}^{2}}{2m}\!+\!\frac{1}{2}m\omega^{2}\bm{r}^{2}\!-\!
 \omega\bm{\sigma}\cdot\bm{L}\right)\!\phi(\bm{r})\!=\!\left(\frac{E^{2}\!-\!m^{2}\!+\!
 3m\omega}{2m}\right)\!\phi(\bm{r}).
\label{SchrodEqn0}
\end{equation} 
This is---indeed---the differential equation of a nonrelativistic, isotropic harmonic oscillator of frequency 
$\omega$ with an added spin-orbit term and an effective nonrelativistic energy
\begin{equation} 
 E_{{}_{\!N\!R}} \equiv \left(\frac{E^{2}\!-\!m^{2}\!+\!3m\omega}{2m}\right).
 \label{ENR0}
\end{equation}
Although the above Schr\"odinger-like equation has been occasionally referred to as the 
\emph{non-relativistic limit} of the Dirac oscillator\,\cite{Moshinsky:1989,Quesne:1990}, we
should underscore that the physical content of Eqs.\,(\ref{DiracEqn1a}) and (\ref{SchrodEqn0}) 
is identical to the one displayed in the original coupled set of Eqs.(\ref{DiracEqn0}). That is,
no approximations have been made and no limits have been taken.

To compute the spectrum of the Dirac oscillator one uses the fact that the spin-spherical harmonics 
are eigenstates of the spin-orbit operator\,\cite{Sakurai:1967}:
\begin{equation} 
 (\bm{\sigma}\cdot\bm{L})|\kappa m\rangle = -(1+\kappa)|\kappa m\rangle.
\end{equation}
The positive energy spectrum of the Dirac oscillator is now readily obtained by enforcing the 
following equality:
\begin{equation} 
 E_{{}_{\!N\!R}}-(1+\kappa)\omega=\left(2n+l+\frac{3}{2}\right)\omega,
 \label{ENR1}
\end{equation}
where the right-hand side of the equation is the energy of the conventional (i.e., without spin orbit) 
nonrelativistic harmonic oscillator. One obtains,
\begin{align}
\frac{E^{2} - m^{2}}{2m} = 
 \begin{cases}
      \ 2 n\omega & \text{if } \kappa < 0, \\        
      \left(2n+2l+1\right) \omega & \text{if } \kappa > 0.
      \end{cases}
 \label{PosEspectrum}     
\end{align} 
This is a very peculiar energy spectrum with energies and degeneracies quite different from those 
of the ordinary harmonic oscillator. For example, for positive values of $\kappa$, i.e., $l\!=\!\kappa\!>\!0$, 
the penalty for adding nodes to the wave function is as costly as increasing the angular-momentum
 barrier. This is unlike the ordinary harmonic oscillator where nodes are twice as costly; see 
 right-hand side of Eq.(\ref{ENR1}). As such, the degeneracy pattern of the Dirac oscillator for 
 $\kappa\!>\!0$ is closer to the hydrogen atom than to the ordinary oscillator. Even more peculiar 
 is the $\kappa\!<\!0$ case ($s^{1/2}$, $p^{3/2}$, $d^{5/2}$, $\ldots$) where the energy depends 
only on the number of nodes $n$ and not on $\kappa$ (or the orbital angular-momentum quantum 
number). That is, for a fixed number of nodes, the Dirac oscillator displays an \emph{infinite degeneracy} 
for $\kappa\!<\!0$.  

Having computed the energy spectrum of the Dirac oscillator, we can now display the associated 
eigenvectors; for more details see the appendix. In particular, the positive energy ($E\!>\!0$) 
solutions of the Dirac oscillator are 
\begin{equation}
\psi_{E\kappa m}(\bm{r})\!=\!
\left(\begin{array}{l}
 \phantom{+i\zeta_{\kappa}}\displaystyle{\sqrt{\frac{E\!+\!m}{2E}}\,R_{nl}(r)}\left|+\kappa m\right\rangle 
 \vspace{5pt} \\
  +i\zeta_{\kappa}\,\displaystyle{\sqrt{\frac{E\!-\!m}{2E}}\,R_{n'l'}(r)}\!\left|-\kappa m\right\rangle
\end{array}\right),
\label{PositiveE}
\end{equation}
while the negative energy ($E\!<\!0$) solutions are given by
\begin{equation}
\psi_{E\kappa m}(\bm{r})\!=\!
\left(\begin{array}{l}
 \phantom{-i\zeta_{\kappa}}\displaystyle{\sqrt{\frac{|E|\!-\!m}{2|E|}}\,R_{nl}(r)}\left|+\kappa m\right\rangle 
 \vspace{5pt} \\
  -i\zeta_{\kappa}\,\displaystyle{\sqrt{\frac{|E|\!+\!m}{2|E|}}\,R_{n'l'}(r)}\!\left|-\kappa m\right\rangle
\end{array}\right).
\label{NegativeE}
\end{equation}
As shown in the appendix, $R_{nl}(r)$ are the radial solutions of the ordinary harmonic oscillator,
$\zeta_{\kappa}\!=\!{\rm sgn}(\kappa)$, and the indices describing the upper and lower components 
are related as follows:
\begin{align}
n' = 
 \begin{cases}
      n & \text{if } \kappa > 0, \\        
      n\!-\!1 & \text{if } \kappa < 0,
      \end{cases} 
      \quad
l' = 
 \begin{cases}
      l\!-\!1 & \text{if } \kappa > 0, \\        
      l\!+\!1 & \text{if } \kappa < 0.
      \end{cases}
 \label{Primes}     
\end{align}       
In particular, note that for $n\!=\!0$ and $\kappa\!<\!0$, one obtains $E\!=\!m$, so the entire
lower component vanishes. The solutions to the Dirac oscillator problem are both intuitive 
and elegant. Given the relation between the number of nodes and the orbital angular momentum 
dictated by the Dirac equation, each component satisfies a Schr\"odinger-like equation 
supplemented by a strong spin-orbit term. As noted earlier, the upper and lower components
have different intrinsic parities, namely, $l'\!=\!l\!\pm\!1$, as a consequence that the orbital 
angular momentum is no longer a good quantum number in the relativistic framework, even
when the potentials are spherically symmetric\,\cite{Sakurai:1967}.

\subsection{\label{subsec:onrmf}Dirac Oscillator Basis: \\ Applications to Covariant DFT}

The previous two sections introduced the Dirac Hamiltonian for a particular version of a covariant 
energy density functional and the Dirac oscillator basis that will be used to create its matrix representation. 
Although we are only interested in the positive energy sector of the DFT Hamiltonian, one must underscore
that the positive energy sector of the Dirac oscillator by itself is not complete, so care must be 
taken to also include the negative energy sector in the construction of the matrix. Considering the 
entire spectrum of the Dirac oscillator, it is convenient to denote its eigenstates as 
$|s n \kappa m\rangle$, with $s$ the sign of the energy.

We now proceed to compute matrix elements of the Hamiltonian given in Eq.(\ref{eq: dirac hamiltonian})
in the Dirac oscillator basis. Before doing so, we rewrite the Hamiltonian by adding and subtracting the
linear term in $\bm{r}$ introduced in Eq.(\ref{eq:DirOscH}). That is, 
$\hat{H}_{\rm DFT}\!\equiv \!\hat{H}_{0}\!+\!\hat{H}_{1}$, where
\begin{subequations}
\begin{align}
  \hat{H}_{0} & =  \bm{\alpha}\!\cdot\!\big(\bm{p}\!-\!im\omega \beta \bm{r}\big)+ \beta m, \\
  \hat{H}_{1} & =  V(r) + \beta S(r) -  im\omega \beta \bm{\alpha}\cdot\bm{r}.
\label{HDFT}                     
\end{align}
\end{subequations}
For a spherically symmetric problem as the one given above, $\kappa$ and $m$ are good quantum numbers,
but the matrix elements are independent of $m$. Thus, one can diagonalize $\hat{H}_{\rm DFT}$ within each 
$\kappa$ block. For an axially-symmetric problem $\kappa$ is no longer a good quantum number so one must 
diagonalize $\hat{H}_{\rm DFT}$ within each individual $m$ block. For this case, matrix diagonalization is 
more efficient than the Runge-Kutta algorithm and this advantage will be explored in a forthcoming work.

By construction, $\hat{H}_{0}$ is diagonal in the Dirac oscillator basis:
\begin{equation}
 \left\langle s^{\prime}n^{\prime}\kappa^{\prime}m^{\prime}\left|\hat{H}_{0}\right|sn\kappa m\right\rangle
  =E_{sn\kappa}\delta_{ss^{\prime}}\delta_{nn^{\prime}}\delta_{\kappa\kappa^{\prime}}\delta_{mm^{\prime}},
\end{equation}
where $E_{sn\kappa}$ is the eigenvalue corresponding to the Dirac oscillator eigenstate $|s n \kappa m\rangle$.
In turn, for the $\hat{H}_{1}$ part of the Hamiltonian that is diagonal in $\kappa$ and $m$ we obtain
\begin{equation}
 \left\langle s^{\prime}n^{\prime}\kappa^{\prime}m^{\prime}\left|\hat{H}_{1}\right|sn\kappa m\right\rangle\!=\!
 \left\langle s^{\prime}n^{\prime}\kappa m \left|\hat{H}_{1}\right|sn\kappa m\right\rangle 
 \delta_{\kappa\kappa^{\prime}}\delta_{mm^{\prime}}
\end{equation}
where
\begin{align}
  \left\langle s^{\prime}n^{\prime}\kappa m \left|\hat{H}_{1}\right|sn\kappa m\right\rangle  & = \nonumber \\
     \int_{0}^{\infty} \Bigg\{V(r) &\Big(g_{\alpha^{\prime}}(r)g_{\alpha}(r)\!+\!f_{\alpha^{\prime}}(r)f_{\alpha}(r)\Big) + \nonumber \\
                                       S(r) & \Big( g_{\alpha^{\prime}}(r)g_{\alpha}(r)\!-\!f_{\alpha^{\prime}}(r)f_{\alpha}(r)\Big) - \nonumber \\
                        (m\omega r) & \Big( g_{\alpha^{\prime}}(r)f_{\alpha}(r)\!+\!f_{\alpha^{\prime}}(r)g_{\alpha}(r)\Big)\Bigg\} dr,
\end{align}
and we have used the short-hand notation $\alpha\!\equiv\!sn\kappa$. Here $g_{\alpha}$ and $f_{\alpha}$ are 
the upper and lower components of the radial wave function of the Dirac oscillator introduced in Eq.(\ref{eq:general solution}). 

\bigskip

\section{\label{sec:result}Results}

{\renewcommand{\arraystretch}{1.2}
\begin{table*}[t]
\begin{tabular}{|c|l||c|c|c|c|}
\hline 
Observable & \hspace{10pt} Method  & $^{40}\rm{Ca}$ & $^{48}\rm{Ca}$ & $^{132}\rm{Sn}$ & $^{208}\rm{Pb}$\tabularnewline
\hline 
\hline 
$B/A$\,(MeV) & Runge-Kutta & -8.538 & -8.584 & -8.339 & -7.889\tabularnewline
                                    & Dirac-Oscillator & -8.539 & -8.585 & -8.339 & -7.888\tabularnewline 
$R_{\rm skin}$\,(fm)    & Runge-Kutta & -0.0513 & 0.1973 & 0.2709 & 0.2069\tabularnewline
                                    & Dirac-Oscillator & -0.0515 & 0.1971 & 0.2712 & 0.2070\tabularnewline
\hline 
\end{tabular}
\caption{
Binding energy per nucleon and neutron skin thickness of $^{40}\text{Ca}$, $^{48}\rm{Ca}$, $^{132}\rm{Sn}$, 
and $^{208}\rm{Pb}$ as predicted by the FSUGold model\,\cite{Todd-Rutel:2005fa}. Self-consistent calculations
were performed using both the Runge-Kutta method and the Dirac oscillator basis.}
\label{table:be}
\end{table*}

\begin{figure*}[ht]
 \includegraphics[width=1.75\columnwidth]{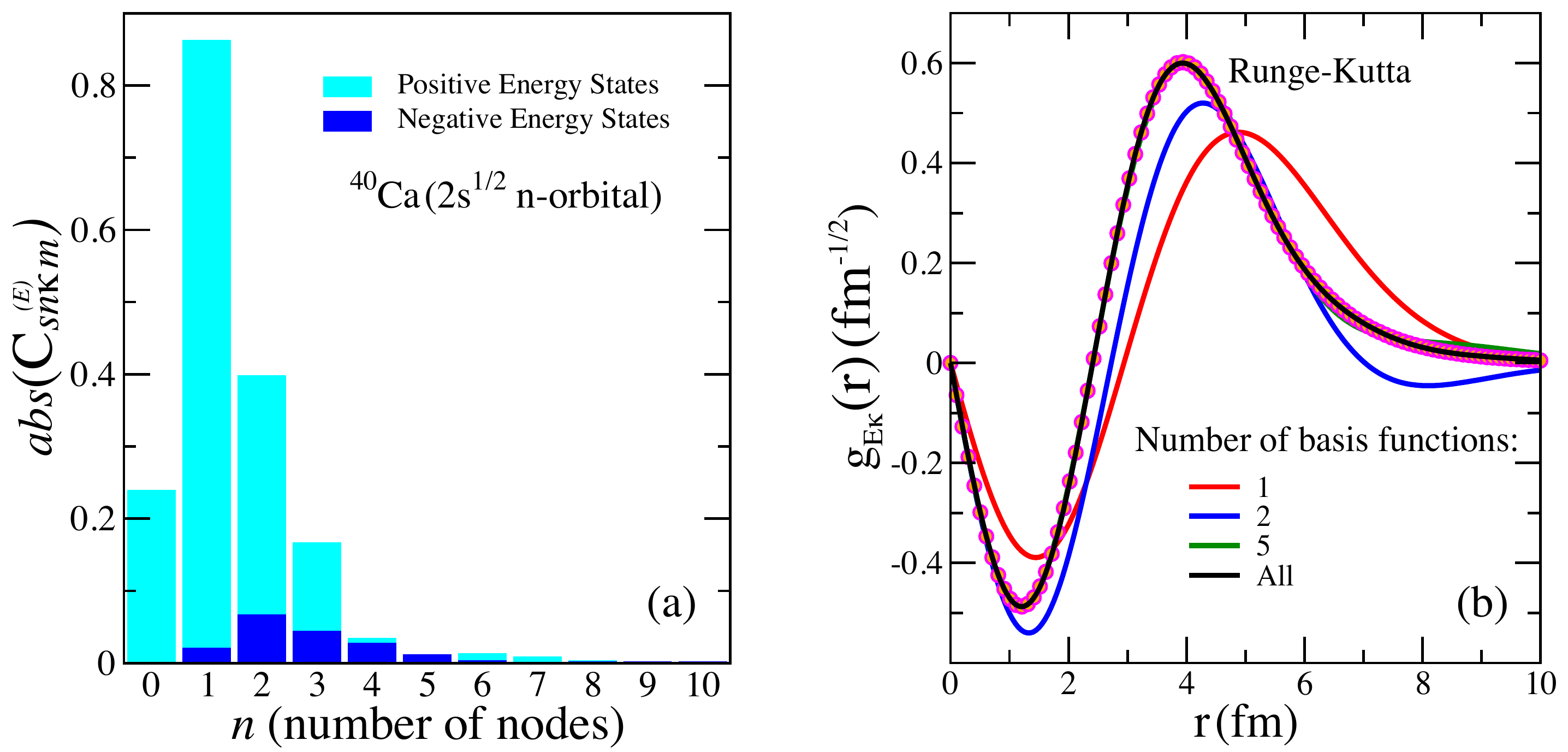}
\caption{(a) Absolute value of the projections (or amplitudes) of the $2s^{\frac{1}{2}}$ ($\kappa\!=\!-1$)
neutron orbital in $^{40}\text{Ca}$ on to the various Dirac oscillator basis states; see Eq.(\ref{eq:coef}). 
The horizontal axis denotes the number of nodes in the first 11 $\kappa\!=\!-1$ basis states. The light (dark)
blue bars represent the contribution from the positive (negative) energy states. (b) Piecewise reconstruction 
of the upper component of $2s^{\frac{1}{2}}$ ($\kappa\!=\!-1$) neutron orbital in $^{40}\text{Ca}$ by combining 
the various Dirac oscillator basis states in order of importance. For example, the red line includes the most 
important basis state (i.e., the $n\!=\!1$ positive energy state). In turn, the blue line includes the two most 
important states and so on. The black line includes the contribution of all basis states used in the diagonalization
procedure. Finally, the curve depicted with the magenta circles was obtained from using the Runge-Kutta
algorithm.} 
\label{Fig1}
\end{figure*}

The main purpose of this section is to test the reliability of the Dirac oscillator basis and to discuss the new insights
that emerge from such an approach. Although we start with scalar and vector potentials that are spherically symmetric, 
we will discuss how to construct the Hamiltonian matrix for the more general case of an axially-symmetric problem 
where the $z$-component of the angular momentum $m$ remains the only good quantum number. One 
starts by dividing the Hamiltonian matrix $H_{\rm DFT}$ into blocks, each with a well defined value of $m$. 
For a given value of $m$, the minimum value of the total angular momentum is given by $j_{\rm min}\!=\!|m|$, which 
in turn implies a minimum value of $|\kappa_{\rm min}|\!=\!j_{\rm min}\!+\!1/2\!=\!|m|+1/2$. The maximum value of 
$\kappa$ is set by the single-particle orbital with the largest value of the total angular momentum. Given that in this 
paper we focus on the four doubly-magic nuclei $^{40}\text{Ca}$, $^{48}\rm{Ca}$, $^{132}\rm{Sn}$, and $^{208}\rm{Pb}$, 
the occupied single-particle orbital with the largest value of the total angular momentum is the $i^{13/2}$ ($\kappa\!=\!-7$) 
neutron orbital in $^{208}\rm{Pb}$. Hence, we set $|\kappa_{\rm max}|\!=\!10$, a value that was found to yield 
well-converged results. Regardless of whether the problem involves spherical symmetry or not, one must also specify 
the maximum value of $n$. As one increases the number of nodes $n$, one can account 
for higher momentum components in the wave function. Although for the most extreme case of $^{208}\rm{Pb}$ no 
single-particle orbital displays more than two nodes, we selected $n_{\rm max}\!=\!10$. The results improve 
very rapidly with increasing $n_{\rm max}$ and are fully converged by $n_{\rm max}\!=\!10$. Note that the range of 
values adopted for $n$ and $\kappa$ are used for both the positive- and negative-energy sector. Also note that
if the entire Hamiltonian is spherically symmetric, one ``discovers" that states with different values of $\kappa$ do
not mix. 

Having identified the Dirac oscillator states that will be used to build the Hamiltonian matrix, one must select the value 
of the oscillator frequency $\omega$, or equivalently the oscillator length parameter $b\!\equiv\!1/\!\sqrt{m\omega}$. 
Although in principle one could optimize the diagonalization by selecting the parameter variationally, for our purposes 
it was sufficient to fix $b$ by demanding good agreement with the lowest $s^{1/2}$ proton orbital in ${}^{208}\rm{Pb}$. 
By doing so, the value of the oscillator parameter was fixed to $b\!=\!2.4\,{\rm fm}$. Although an optimal value of $b$ 
reduces the number of basis states required to reproduce the entire spectrum, the adopted value of $b$---together with 
the range of values chosen for $n$ and $\kappa$---produced stable results for all nuclei under consideration. Finally, to 
test the reliability of the method we selected the FSUGold model introduced in Ref.\,\cite{Todd-Rutel:2005fa}.

Predictions for the binding energy per nucleon and the neutron skin thickness of the doubly-magic nuclei 
$^{40}\text{Ca}$, $^{48}\rm{Ca}$, $^{132}\rm{Sn}$, and $^{208}\rm{Pb}$ are displayed in Table.\,\ref{table:be} 
using both the Runge-Kutta algorithm as well as the Dirac oscillator basis. The neutron skin thickness, a 
sensitive isovector observable, is defined as the difference between the neutron and proton root-mean-square 
radii. Evidently, the agreement between the two methods is excellent---even when the small neutron skin 
thickness emerges from the difference of two radii of similar size. 

Besides the fact that the Dirac oscillator method can be easily generalized to the study of axially-deformed 
nuclei, the method also provides valuable insights. For example, how are the eigenstates of 
$\hat{H}_{\rm DFT}$  expressed as a linear combination of the Dirac oscillator states? To answer this question 
we select the $2s^{1/2}$ neutron orbital in $^{40}\text{Ca}$ as an example; that is, the orbital with one 
interior node. In this case spherical symmetry is still preserved so $\kappa\!=\!-1$ emerges as a good quantum 
number. The corresponding eigenstate may be expressed in terms of the Dirac oscillator basis as follows:
\begin{equation}
 |\psi_{E\kappa m}\rangle = \sum_{sn} C_{sn\kappa m}^{(E)} |s n \kappa m\rangle,
 \label{eq:coef}
\end{equation}
where the associated amplitudes $C_{sn\kappa m}^{(E)}$ that emerge from the diagonalization procedure 
satisfy the normalization condition 
\begin{equation}
  \sum_{sn} \left|C_{sn\kappa m}^{(E)}\right|^{2} = 1.
 \label{normal}
\end{equation}

We display on the left-hand panel of Fig.\,\ref{Fig1} the absolute value of the amplitudes 
$C_{sn\kappa m}^{(E)}$ for the $2s^{1/2}$ neutron orbital in $^{40}\text{Ca}$ as a 
function of the number of nodes of each individual basis state $|s, n,\kappa\!=\!-1, m\!=\!0\rangle$.
However, since the positive energy states by themselves do not form a complete basis, we depict 
with the light-blue bars the projections (or amplitudes) into the positive energy states and with 
dark-blue bars the corresponding projections into the negative energy states. As expected, the 
largest amplitude is carried by the positive energy state having one interior node. The contribution 
from this one basis state to the entire upper component of the wave function may be seen on the 
right-hand panel of Fig.\,\ref{Fig1}. The $n\!=\!2, 0, 3$ are the next most important basis states,
respectively. Yet, the next most important contribution after that comes from the $n\!=\!2$ negative 
energy state. Indeed, with only 5 basis states one can accurately capture the shape of the exact 
wave function; see Fig.\,\ref{Fig1}(b). Although small, the contribution from the negative energy sector 
is vital to accurately reproduce the entire wave function, which is displayed with the black solid line 
(labeled ``All") in Fig.\,\ref{Fig1}(b). The curve depicted with the red circles represents the exact 
solution obtained using the Runge-Kutta method and it is clearly indistinguishable from the black line. 
This simple, yet illustrative example, confirms that the Dirac oscillator basis is both efficient and 
insightful for the solution of relativistic nuclear-structure problems. 

\begin{figure*}[ht]
 \includegraphics[width=1.75\columnwidth]{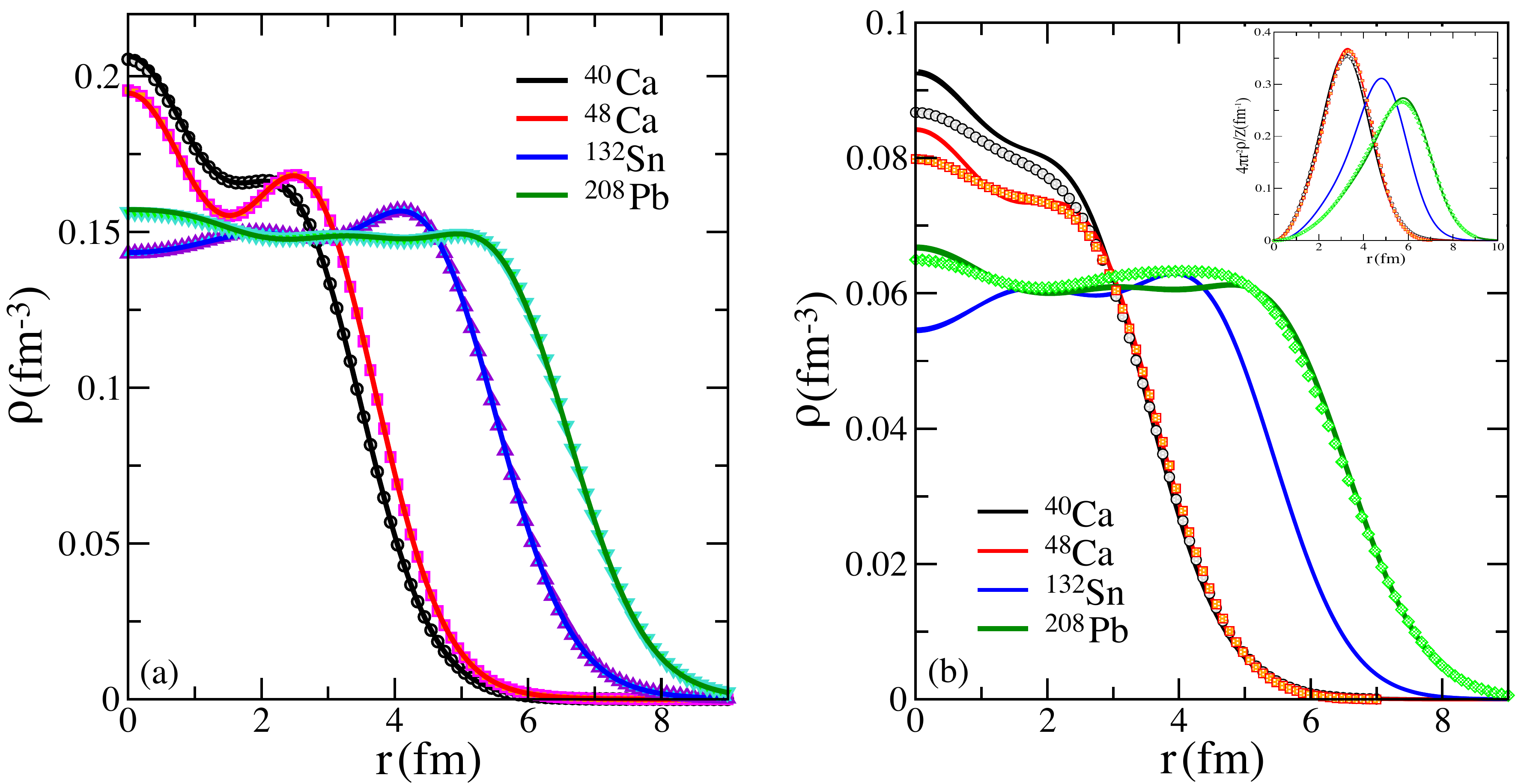}
\caption{(a) Baryon (neutron-plus-proton) densities of $^{40}\text{Ca}$, $^{48}\rm{Ca}$, $^{132}\rm{Sn}$, 
and $^{208}\rm{Pb}$ as predicted by the FSUGold model\,\cite{Todd-Rutel:2005fa}. Results from the
Dirac oscillator method are denoted with the solid lines whereas those obtained from the Runge-Kutta 
algorithm are depicted with the various symbols. (b) Same as in panel (a) but now for the charge density.
The nearly identical predictions from the Dirac oscillator and Runge-Kutta methods are plotted jointly as 
solid lines whereas the experimental results\,\cite{DeJager:1987qc} are depicted with the various symbols. 
Note that at present there is no measurement of the charge distribution of $^{132}\rm{Sn}$, although 
see\,\cite{Suda:2009zz,Tsukada:2017llu}. Finally, the inset displays the charge density multiplied by a
suitable phase-space factor that makes the integral under the curve identically equal to one for all nuclei.}
\label{Fig2}
\end{figure*}

We close this section by displaying in Fig.\,\ref{Fig2} the baryon (neutron-plus-proton) density as well 
as the charge density as predicted by the FSUGold model\,\cite{Todd-Rutel:2005fa} for $^{40}\text{Ca}$, 
$^{48}\rm{Ca}$, $^{132}\rm{Sn}$, and $^{208}\rm{Pb}$. On the left-hand panel we highlight the excellent 
agreement between the Runge-Kutta and Dirac oscillator methods in predicting the baryon density---a 
ground-state property that is sensitive to all  nucleon orbitals. Given the excellent agreement between 
the two methods, we display in Fig.\,\ref{Fig2}(b) their predictions (combined in a single solid line) 
alongside the experimental charge density (depicted with symbols) for $^{40}\text{Ca}$, $^{48}\rm{Ca}$, and 
$^{208}\rm{Pb}$\,\cite{DeJager:1987qc}. Note that at present the charge density of $^{132}\rm{Sn}$
is unknown, although enormous progress is being made in the development of electron-scattering 
techniques for the measurement of the charge density of short-lived isotopes\,\cite{Suda:2009zz,
Tsukada:2017llu}. Although the charge radius of several closed-shell nuclei was used in the 
calibration of the FSUGold functional, a common deficiency of models of this kind is the poor 
reproduction of the interior density, a behavior that is controlled by the high-momentum components 
of the charge form factor. Perhaps a more reliable comparison between theory and experiment is the 
phase-space weighted charge density displayed in the inset of Fig.\,\ref{Fig2}(b) and defined as,
\begin{equation}
  \widetilde\rho(r) \equiv \frac{1}{Z}4\pi r^{2}\rho(r) 
  \hspace{10pt} {\rm with} \hspace{10pt}
  \int_{0}^{\infty} \hspace{-5pt}\widetilde\rho(r) dr = 1. 
\end{equation}
In particular, $\widetilde\rho(r)$ has been normalized to 1 for all nuclei and its second moment equals 
the mean-square radius of the charge distribution. 

\section {\label{sec:conclusion}Conclusion}

The staple of nuclear structure is the nuclear shell model, a theoretical framework that in its simplest 
form consists of a harmonic oscillator model supplemented by a strong spin-orbit 
interaction\,\cite{Haxel:1949,Mayer:1949}. Although more than seven decades have passed since its 
original formulation, this simplest version of the nuclear shell model continues to be used today---often
as the first step in the development of more sophisticated models. Initially modeled after the Thomas 
term in atomic systems, it was soon realized that: {\sl ``There is no adequate theoretical reason for 
the large observed value of the spin orbit coupling "}\,\cite{Mayer:1950a}. Ultimately, of course, the
complex dynamical origin of the nuclear spin-orbit force hides within QCD. Nevertheless, inspired
by Yukawa's meson theory, early attempts at building a relativistic models of the nuclear force 
considered nucleons interacting by the exchange of isoscalar meson fields of Lorentz scalar and 
vector character\,\cite{Johnson:1955zz,Duerr:1956zz,Miller:1972zza,Walecka:1974qa}. While
successful in many regards\,\cite{Serot:1984ey}, the models also provided a natural explanation for 
the emergence of a strong spin-orbit force. Since then, relativistic nuclear models have been 
augmented and refined---and are now part of a vast arsenal of theoretical tools devoted to the
study of diverse nuclear phenomena. It is in this overall context that we find surprising that the 
Dirac oscillator model of Moshinsky and Szczepaniak\,\cite{Moshinsky:1989}---which may be 
cast in the form of an ordinary oscillator model with a strong spin-orbit term---has remained largely 
unknown to the nuclear physics community. The present contribution aims to remedy this situation.

In this paper we introduced the Dirac oscillator and highlighted some of its remarkable properties. 
Mainly, the fact that the Dirac oscillator can be solved exactly given that both of its upper and lower
components satisfy Schr\"odinger-like equations identical to the conventional harmonic oscillator 
supplemented by a strong spin-orbit interaction. Matrix elements of a mean-field Hamiltonian 
containing strong scalar and (time-like) vector potentials were computed in the Dirac oscillator
basis. Once the matrix elements were obtained, we carried out the entire self-consistent procedure, 
which involved solving Klein-Gordon equations for the meson fields and diagonalizing the mean-field 
Hamiltonian for the nucleon field until convergence was achieved. By selecting a reasonable value 
for the oscillator frequency, or equivalent the oscillator length, relatively modest-size matrices had
to be diagonalized. We note that while the contribution from the negative energy sector of  the Dirac 
oscillator played a relatively minor role, it was by no means negligible. Finally, we compared results
against those obtained using the Runge-Kutta method. In all instances the comparisons between the 
two methods were excellent. Yet, we argue that relative to the Runge-Kutta method, the Dirac oscillator 
method has the distinct advantage that the generalization to problems with broken spherical symmetry 
is fairly straightforward. 

Problems with axial symmetry may play a pivotal role in the refinement of modern energy density 
functionals. The standard fitting protocol of energy density functionals (both covariant and non-relativistic) 
involves calibrating the parameters of the model by invoking genuine physical observables---mostly 
ground-state properties such as binding energies and charge radii. However, these observables are 
insensitive to certain critical aspects of the nuclear dynamics. For example, whereas binding energies 
and charge radii effectively constrain the saturation point  (density and binding energy per nucleon)
of symmetric nuclear matter, they provide little guidance on the incompressibility coefficient. In an effort 
to remedy this deficiency, experimental information on isoscalar giant monopole energies---which are 
strongly correlated to the incompressibility coefficient---are now incorporated into the calibration of
the functionals\,\cite{Chen:2014sca}. However, in an effort to reduce the computational demands, 
giant-monopole energies were computed in a constrained approach, thereby avoiding the need for 
computationally demanding random-phase-approximation (RPA)
calculations\,\cite{Chen:2013tca}. In a constrained approach, giant-monopole energies may be directly 
computed from a ground-state calculation by adding to the Hamiltonian a ``constrained" one-body term 
proportional to the operator responsible for the excitation, namely, $\lambda r^{2}$. Note that for 
monopole excitations the Hamiltonian remains spherically symmetric.  

Whereas binding energies, charge radii, and giant-monopole energies provide stringent constraints 
on the isoscalar sector of the functional, the lack of experimental data on exotic nuclei with a large 
neutron-proton asymmetries has hindered our knowledge of the isovector sector. It has been recognized 
that the neutron skin thickness of neutron rich nuclei such as ${}^{208}$Pb and ${}^{48}$Ca can 
effectively constrain the isovector sector, particularly the density dependence of the nuclear symmetry 
energy, and experimental campaigns are currently underway at JLab to determine the thickness of the
neutron skin in a clean and largely model independent way\,\cite{Abrahamyan:2012gp,Horowitz:2012tj,
CREX:2013}}. The electric dipole polarizability has also been shown to be a strong isovector 
indicator\,\cite{Reinhard:2010wz,Piekarewicz:2012pp,Roca-Maza:2013mla}. 
The electric dipole polarizability is a static observable that is proportional to the inverse energy weighted 
sum $m_{-1}$ of the isovector dipole response. As such, $m_{-1}$ may be computed from a ground-state
calculation by adding a ``constrained" one-body term of the form: $\lambda r Y_{1,0}(\theta,\varphi) \tau_z$. 
As already shown here, matrix elements of operators of the general form $r^{L}Y_{L,0}(\theta,\varphi)$
may be readily evaluated in the Dirac oscillator basis. Hence, in principle one may incorporate experimental 
information on the electric dipole polarizability in the calibration of the functional, by either computing 
$m_{-1}$ in a constrained approach or by computing the entire isovector dipole response in a relativistic 
random phase approximation. Although successful in computing various moments of the monopole and 
quadrupole distribution in a constrained approach, so far we have been unable to obtain stable results 
for the electric dipole polarizability. To date, the only successful approach that we are aware of is the one 
by Yuksel, Marketin, and Paar\,\cite{Yuksel:2019dnp}, in which the calibration of a new ``point coupling" 
model involved computing the entire isovector dipole response.

In summary, we have demonstrated that the Dirac oscillator basis provides an ideal tool for solving
nuclear-structure problems formulated within the framework of covariant density functional theory. 
Indeed, excellent agreement was obtained when compared against results generated using the 
Runge-Kutta method. The Dirac oscillator incorporates two features that have been at the core of 
nuclear structure since its inception: (i) a harmonic oscillator basis supplemented by (ii) a strong 
spin-orbit coupling. 
In this paper we outlined the steps that are necessary to compute matrix elements of an axially 
symmetric relativistic mean-field Hamiltonian in the Dirac oscillator basis. Problems of this kind are 
useful for understanding the structure of deformed nuclei as well as in computing various moments 
of the nuclear response in a constrained approach. Ultimately, an efficient and robust computation 
of these moments could significantly improve the quality of future covariant energy density functionals.

\begin{acknowledgments}
This material is based upon work supported by the U.S. Department of Energy 
Office of Science, Office of Nuclear Physics under Award Number DE-FG02-92ER40750. 
\end{acknowledgments}
\vfill

\section{\label{sec:append}Appendix}
\renewcommand\theequation{A.\arabic{equation}}
\setcounter{equation}{0}
In this section we provide a detailed derivation of the energies and corresponding 
eigenfunctions of the relativistic Dirac oscillator. Given the spherical symmetry of the 
potential, we start by writing the 4-component solution as follows\,\cite{Sakurai:1967}:
\begin{equation}
\psi_{E\kappa m}(\bm{r})\!=\!
\left(\begin{array}{c}
\phi_{E\kappa m}(\bm{r})\\
\chi_{E\kappa m}(\bm{r})
\end{array}\right)
\!=\!
\left(\begin{array}{c}
\phantom{i}g_{E\kappa}(r)\left|+\kappa m\right\rangle \\
if_{E\kappa}(r)\left|-\kappa m\right\rangle
\end{array}\right),
\label{Appx1}
\end{equation}
where both $g_{E\kappa}(r)$ and $f_{E\kappa}(r)$ are real functions of $r$ and the spin-spherical 
harmonics $|\kappa m\rangle$ have been defined in Eq.\,(\ref{DefKappa}). As shown in 
Eq.\,(\ref{SchrodEqn0}), the upper component of the Dirac equation satisfies a Schr\"odinger-like 
equation supplemented by a strong spin-orbit term:
\begin{equation} 
 \left(\frac{\bm{p}^{2}}{2m}\!+\!\frac{1}{2}m\omega^{2}\bm{r}^{2}\!-\!
 \omega\bm{\sigma}\cdot\bm{L}\right)\!\phi(\bm{r})\!=\!\left(\frac{E^{2}\!-\!m^{2}\!+\!
 3m\omega}{2m}\right)\!\phi(\bm{r}).
\label{Appx2}
\end{equation} 
Because of the appearance of the spin-orbit interaction, the energies of the Dirac oscillator 
depend strongly on the generalized angular momentum quantum number $\kappa$. That is, 
\begin{align}
\frac{E^{2} - m^{2}}{2m} = 
 \begin{cases}
      \ 2 n\omega & \text{if } \kappa < 0, \\        
      \left(2n+2l+1\right) \omega & \text{if } \kappa > 0.
      \end{cases}
 \label{Appx3}     
\end{align} 
In particular, for $\kappa\!<\!0$ the energies are independent of any angular momentum quantum 
number (i.e., $l$ and $j$) so the spectrum---for a fixed value of $n$---displays an infinite degeneracy. 
Although more reminiscent of the ordinary oscillator, the energies for $\kappa\!>\!0$ weigh equally 
the number of nodes $n$ as the orbital angular momentum quantum number $l$; recall that in the 
ordinary case nodes cost twice as much energy as $l$. 

At this point it is useful to invoke some well known results from the ordinary harmonic oscillator. 
For example, the radial solution of the isotropic, three dimensional harmonic oscillator is 
\begin{equation}
 R_{nl}(x) = A_{nl}\,x^{l} {\cal M}\!\left(-n,l\!+\!\frac{3}{2},x^{2}\right)\expo^{-x^{2}/2},
 \label{Appx4}
\end{equation}
where $n$ is the number of interior nodes, $l$ is the orbital angular momentum, 
$x\!\equiv\!r/b$ is the dimensionless radial distance measured in units of the 
oscillator-length parameter $b\!=\!1/\sqrt{m\omega}$, and $A_{nl}$ is the 
normalization constant given by
\begin{equation}
 A_{nl} = \sqrt{\frac{2\,\Gamma(n\!+\!l\!+\!3/2)}{(n!)\,\Gamma^{2}(l\!+\!3/2)}}.
 \label{Appx5}
\end{equation}
Besides the characteristic Gaussian falloff, the radial dependence of $R_{nl}(r)$ is 
contained in Kummer's function\,\cite{Abramowitz:1972}:
\begin{subequations}
\begin{align}
  & {\cal M}(\alpha,\beta,z) = \sum_{m=0}^{\infty} \frac{(\alpha)_{m}}{(\beta)_{m}}\frac{z^{m}}{m!}\,, \\
  {\rm with} \hspace{5pt}  & (\alpha)_{m} \equiv \alpha(\alpha+1)(\alpha+2)\ldots(\alpha+m-1).
 \label{Appx6}
\end{align}
\end{subequations}
Although in principle the above sum extends up to infinity, in practice the sum is truncated 
after $n\!+\!1$ terms, making ${\cal M}(-n,\alpha,\beta)$ a polynomial of degree $n$; note 
that $(\alpha)_{0}\!\equiv\!1$. In this way, we can write the upper component of the radial 
solution of the Dirac equation simply as
\begin{equation}
 g_{E\kappa}(r) = R_{nl}(x\!=\!r/b),
  \label{Appx7}
\end{equation}
where the energy is given in Eq.\,(\ref{Appx3}) and $\kappa$ is related to $l$ via Eq.\,(\ref{DefKappa}).
Having solved for the upper component, the lower component may be obtained by differentiation. That
is, using Eq.\,(\ref{DiracEqn1a}) one obtains
\begin{align}
 & \chi(\bm{r})  = \frac{(\bm{\sigma}\cdot\bm{\pi}_{-})}{(E+m)}\phi(\bm{r}) =
 \frac{\bm{\sigma}\cdot\left({\bf p}-im\omega{\bf r}\right)}{(E+m)}\phi(\bm{r}) \nonumber \\
                    & = \frac{i}{(E\!+\!m)}\left[\frac{\partial}{\partial r}\!+\!\frac{(\kappa\!+\!1)}{r}\!+\!m\omega r\right]
                       \!g_{E\kappa}(r)\left| -\kappa m\rangle\right. .
 \label{Appx8} 
\end{align}
Comparing this expression to Eq.(\ref{Appx1}) we conclude that
\begin{align}
 f_{E\kappa}(r) = &\left(\frac{1}{E+m}\right)
 \left[\frac{d}{dr} \!+\! \frac{(\kappa\!+\!1)}{r} \!+\! m\omega r\right]g_{E\kappa}(r) \nonumber\\
= &\left(\frac{b^{-1}}{E+m}\right)
 \left[\frac{d}{dx} \!+\! \frac{(\kappa\!+\!1)}{x} \!+\! x\right]R_{nl}(x).
 \label{Appx9} 
\end{align}
Now, by direct differentiation of Eq.\,(\ref{Appx4}) one obtains
\begin{align}
 & \left[\frac{d}{dx} \!+\! \frac{(\kappa\!+\!1)}{x} \!+\! x\right]R_{nl}(x) = 
  \nonumber \\
 A_{nl}\,x^{l} & \left[2xM^{\prime}(\alpha,\beta,x^{2})\!+\!
 \frac{(l\!+\!\kappa\!+\!1)}{x}M(\alpha,\beta,x^{2})\right]\expo^{-x^{2}/2},
 \label{Appx10} 
\end{align} 
where $\alpha\!=\!-n$, $\beta\!=\!l+3/2$, and the ``prime" indicates differentiation
with respect to the argument (i.e., $x^{2}$). At this point one must distinguish between 
$\kappa\!<\!0$ and $\kappa\!>\!0$ and for this we use the following two useful relations
involving Kummer's functions\,\cite{Abramowitz:1972}:
\begin{subequations}
\begin{align}
 \frac{dM(\alpha,\beta,z)}{dz}  & \!=\! \frac{\alpha}{\beta}M(\alpha\!+\!1,\beta\!+\!1,z), \\
 z\frac{dM(\alpha,\beta,z)}{dz} & \!=\! (\beta\!-\!1)
 \left[M(\alpha,\beta\!-\!1,z)\!-\!M(\alpha,\beta,z)\right], 
 \label{Appx11} 
\end{align} 
\end{subequations}
where the first identity is used for $\kappa\!<\!0$  and the second one for $\kappa\!>\!0$.
Using these relations one obtains simple and illuminating expressions for the lower 
component of the Dirac oscillator
\begin{align}
 f_{E\kappa}(r) & = \left(\frac{b^{-1}}{E+m}\right)
  \left[\frac{d}{dx} \!+\! \frac{(\kappa\!+\!1)}{x} \!+\! x\right]R_{nl}(x) \nonumber \\
  & = \zeta_{\kappa}\frac{\sqrt{E^{2}-m^{2}}}{E+m}\,R_{n'l'}(x),
 \label{Appx12} 
\end{align} 
where $\zeta_{\kappa}\!=\!{\rm sgn}(\kappa)$, and $n'$ and $l'$ are defined as 
\begin{align}
n' = 
 \begin{cases}
      n & \text{if } \kappa > 0, \\        
      n\!-\!1 & \text{if } \kappa < 0,
      \end{cases} 
      \quad
l' = 
 \begin{cases}
      l\!-\!1 & \text{if } \kappa > 0, \\        
      l\!+\!1 & \text{if } \kappa < 0.
      \end{cases}
 \label{Appx13}     
\end{align}       
This illustrates the well known fact that the upper and lower components of the 
Dirac wave function have different intrinsic parities, namely, $l'\!=\!l\!\pm\!1$.
That is, in the relativistic framework the orbital angular momentum is no longer
a good quantum number. 

We can now proceed to write the properly normalized 
eigenstates of the Dirac oscillator. For the positive energy ($E\!>\!0$) sector one 
obtains
\begin{equation}
\psi_{E\kappa m}(\bm{r})\!=\!
\left(\begin{array}{l}
 \phantom{+i\zeta_{\kappa}}\displaystyle{\sqrt{\frac{E\!+\!m}{2E}}\,R_{nl}(r)}\left|+\kappa m\right\rangle 
 \vspace{5pt} \\
  +i\zeta_{\kappa}\,\displaystyle{\sqrt{\frac{E\!-\!m}{2E}}\,R_{n'l'}(r)}\!\left|-\kappa m\right\rangle
\end{array}\right).
\label{Appx14}
\end{equation}
In turn, eigenstates of the Dirac oscillator with negative energy ($E\!<\!0$) are given by
\begin{equation}
\psi_{E\kappa m}(\bm{r})\!=\!
\left(\begin{array}{l}
 \phantom{-i\zeta_{\kappa}}\displaystyle{\sqrt{\frac{|E|\!-\!m}{2|E|}}\,R_{nl}(r)}\left|+\kappa m\right\rangle 
 \vspace{5pt} \\
  -i\zeta_{\kappa}\,\displaystyle{\sqrt{\frac{|E|\!+\!m}{2|E|}}\,R_{n'l'}(r)}\!\left|-\kappa m\right\rangle
\end{array}\right).
\label{Appx14}
\end{equation}

Such a simple expression is reminiscent of the plane-wave (i.e., free) solutions of the Dirac equation, 
which for positive energy are 
\begin{subequations}
\begin{align}
 g_{E\kappa}(r) & = j_{l}(kr) ,\\
 f_{E\kappa}(r) & = \pm \sqrt{\frac{E-m}{E+m}}\,j_{l'}(kr), 
 \label{Appx15}     
\end{align} 
\end{subequations}
where $k\!=\!\sqrt{E^{2}\!-\!m^{2}}$, and $j_{l}(x)$ are $j_{l'}(x)$ are spherical Bessel 
functions. The remarkable feature of Eq.\,(\ref{Appx14}) is that such a compact 
expression emerges even in the case of the highly non-trivial Dirac oscillator.
\vfill

\bibliography{./DiracOscillator_Final.bbl} 
\end{document}